# A TOSA/ROSA-Based Optical Transmitter (MTx+)/Transceiver (MTRx+) for High-Energy Physics Experiments

**B. Deng,**[a,b,c] **X. Zhao,**[c] **W. Zhou,**[d] **C. Chen,**[c,d] **D. Gong,**[c] **D. Guo,**[d] **S. Hou,**[e] **K. Jin,**[d]
**C. Liu,**[c] **J. Liu,**[d] **T. Liu,**[c,*] **M. Qi,**[f] **Q. Sun,**[c] **J. Thomas,**[c] **X. Le,**[d] **and J. Ye**[c]

[a] *Department of Electronic Information Engineering, Hubei Polytechnic University,*
  *Huangshi, Hubei 435003, P.R. China*

[b] *Zhongyi Environmental Protection College, Yixing Environmental Protection Science and Technology Industrial Park,*
  *Yixing, Jiangsu 214200, P.R. China*

[c] *Department of Physics, Southern Methodist University,*
  *Dallas, TX 75275, U.S.A*

[d] *Department of Physics, Central China Normal University,*
  *Wuhan, Hubei 430079, P.R. China*

[e] *Institute of Physics, Academia Sinica,*
  *Nangang, Taipei 11529, Taiwan*

[f] *Department of Physics, Nanjing University,*
  *Nanjing, Jiangsu 210000, P.R. China*

  *E-mail*: tliu@mail.smu.edu

ABSTRACT: We present a dual-channel optical transmitter (MTx+)/transceiver (MTRx+) for the front-end readout electronics of high-energy physics experiments. MTx+ utilizes two Transmitter Optical Sub-Assemblies (TOSAs) and MTRx+ utilizes a TOSA and a Receiver Optical Sub-Assemblies (ROSA). Both MTx+ and MTRx+ receive multimode fibers with standard Lucent Connectors (LCs) as the optical interface and can be panel or board mounted to a motherboard with a standard Enhanced Small Form-factor Pluggable (SFP+) connector as the electrical interface. MTx+ and MTRx+ employ a dual-channel Vertical-Cavity Surface-Emitting Laser (VCSEL) driver ASIC called LOCld65, which brings the transmitting data rate up to 14 Gbps per channel. MTx+ and MTRx+ have been tested to survive 4.9 kGy($SiO_2$).

---

[*] Corresponding author.

## Contents

## 1. Introduction

In high-energy physics experiments, the data are usually transmitted from the front-end detector to the back-end control room using optical fibers. Due to the harsh radiation environment, the optical transmitter/transceiver mounted on the front-end detector must meet radiation tolerance requirements [1]. Based on a Transmitter Optical Sub-Assembly (TOSA) and a Receiver Optical Sub-Assemblies (ROSA), a small-form-factor optical transceiver VTRx [2] is developed. Following the same idea, we designed a mid-board miniature dual-channel optical transmitter (MTx) and a similar optical transceiver (MTRx) [3] for the ATLAS Liquid Argon Calorimeter trigger readout upgrade [4, 5]. MTx utilizes two TOSAs and MTRx utilizes a TOSA and a ROSA. Comparing with VTRx, MTx and MTRx limit their height to 6 mm. Contrasting with the optical transceiver VTRx+ [6, 7, 8, 9], which is based on a Vertical-Cavity Surface-Emitting Laser (VCSEL) array, a TOSA/ROSA-based optical transmitter/transceiver avoids the complicated precision optical coupling procedure in the assembly.

MTx and MTRx employ a dual-channel VCSEL driver ASIC called LOCld [10]. LOCld is designed in commercial 250-nm Silicon on Sapphire CMOS technology. Each channel of LOCld operates at a data rate up to 8 Gbps. After the production of LOCld, we designed a dual-channel VCSEL driver LOCld65 [11] in commercial 65-nm CMOS technology. Each channel of LOCld65 operates at a data rate up to 14 Gbps.

Continuing the idea of a TOSA/ROSA-based optical transmitter/transceiver with a small form factor and the 6-mm height, we set out to develop MTx+ and MTRx+ with the following requirements. Firstly, we adopt the electrical connector of an Enhanced Small Form-factor Pluggable (SFP+) module to avoid the fragile board-stacking connector in MTx. Secondly, we comply with standard optical Lucent Connectors (LCs) without any customization, no matter how small that may be. Thirdly, we bring the transmitting data rate to 14 Gbps per fiber. Lastly, we make the module board and front-panel mountable.

## 2. Design of MTx+ and MTRx+

An MTx+ (MTRx+) module consists of two TOSAs (a TOSA and a ROSA), a Printed Circuit Board (PCB), and a mechanical latch. MTx+/MTRx+ uses the same TOSAs (Part No. TTF-1F59-427 produced by Truelight) [12] as MTx. The same GBTIA-embedded ROSA [13] as VTRx and MTRx will be utilized in MTRx+ in the future production, but a commercial ROSA (Part No. HFD6180 produced by Finisar) is employed for prototype testing at present. The height of the TOSA/ROSA limits that of MTx+/MTRx+. Each TOSA or ROSA is attached to the PCB through a printed flexible cable.

The core component on the PCB is LOCld65. LOCld65 contains two separate channels and a shared I$^2$C slave. Each channel includes an input amplifier, four stages of limiting amplifiers, a high-current output driver, and a bias-current generator. The I2C slave is protected from single event upsets (SEUs) with triple-modular redundancy. Each channel of LOCld65 is tested to operate up to 14 Gbps.

The PCB has a custom-defined 20-pin gold-finger connector as the interface of MTx+ with a motherboard. The PCB implements AC coupling to the motherboard and to the TOSAs. For MTRx+, there is no component on the receiver channel of the PCB. The gold fingers on the PCB mate with a standard SFP+ connector on the motherboard. The PCB pictures of MTx+ and MTRx+ are shown in Figure 1.

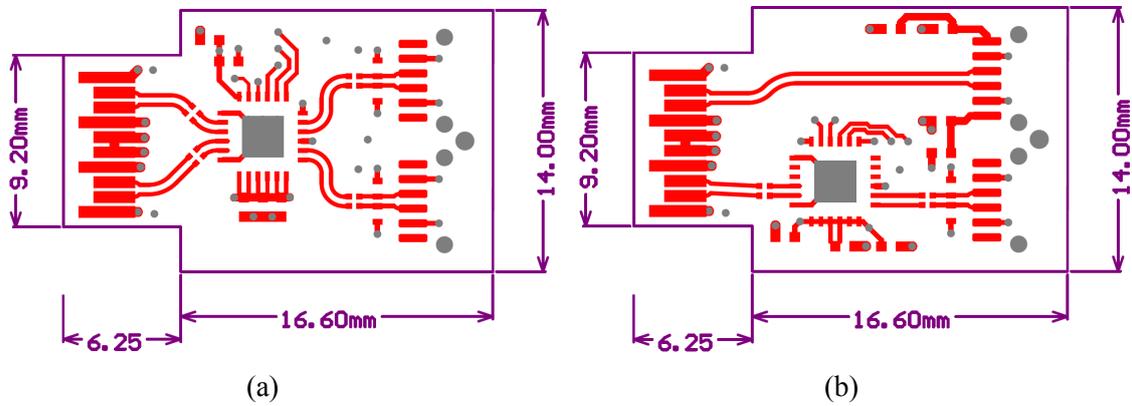

(a)  (b)

**Figure 1.** PCB design of MTx+ (a) and MTRx+ (b).

In MTx+ (or MTRx+), a mechanical latch holds the two LC connectors, two TOSAs (or a TOSA and a ROSA), and the PCB together. When an optical fiber with a standard LC connector is plugged in the latch, the latch holds the ferrule of the fiber in the housing of the TOSA/ROSA and ensures a proper optical coupling between the fiber and the TOSA/ROSA. The bottom view of the latch is shown in Figure 2(a). The bottom view and the top view of an MTx+ module with an optical latch and two TOSAs are shown in Figure 2(b) and Figure 2(c), respectively. In order to minimize the module height, the two LC connectors face back to back. The latch has three mechanical holes, the central one to attach the latch to the PCB, the two side ones to fix the latch to the motherboard. The prototype latch is made in injection molding from the same material of the body of the TOSA/ROSA.

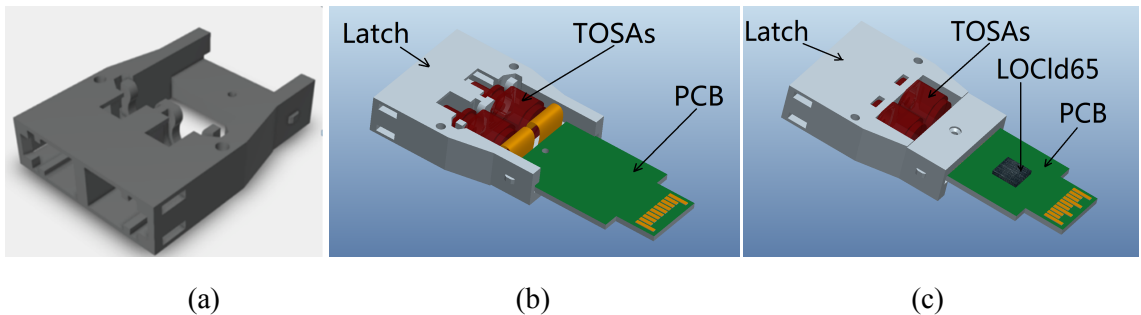

(a)　　　　　　　　　　　(b)　　　　　　　　　　　(c)

**Figure 2.** Bottom view of a latch (a), bottom view (b) and top view (c) of an MTx+ module.

A picture of two latches (a top view and a bottom view) and two metal cages (a top view and a bottom view) is shown in Figure 3(a). A picture of an MTx+ module is shown in Figure 3(b). The size of MTx+/MTRx+ is 43.5 mm (length) × 19 mm (width) × 5.9 mm (height). An MTx+/MTRx+ module is 6 mm when the metal cage and the LC connector are counted. As a comparison, the size of MTx/MTRx is 45 mm (length) × 15 mm (width) × 6 mm (height).

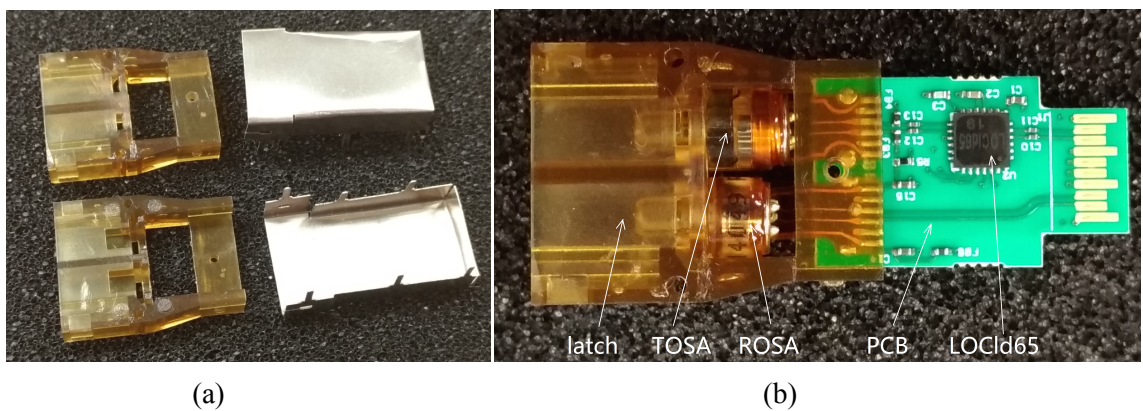

(a)　　　　　　　　　　　　　　　(b)

**Figure 3.** Pictures of latches and cages (a) and an MTRx+ module (b).

MTx+/MTRx+ can be either front-panel mounted or board mounted. Figure 4(a) shows the pictures of a panel-mounted module and Figure 4(b) displays a board-mounted (b) module onto the motherboard. In the case of panel mounted MTx+, as shown in Figure 4(a), the front-panel position is marked as the two red lines. A metal cage attaches the module to the motherboard like in the case of SFP+. Prototype cages have been fabricated. In the case of board-mounted MTx+, two screws fix the latch to the motherboard.

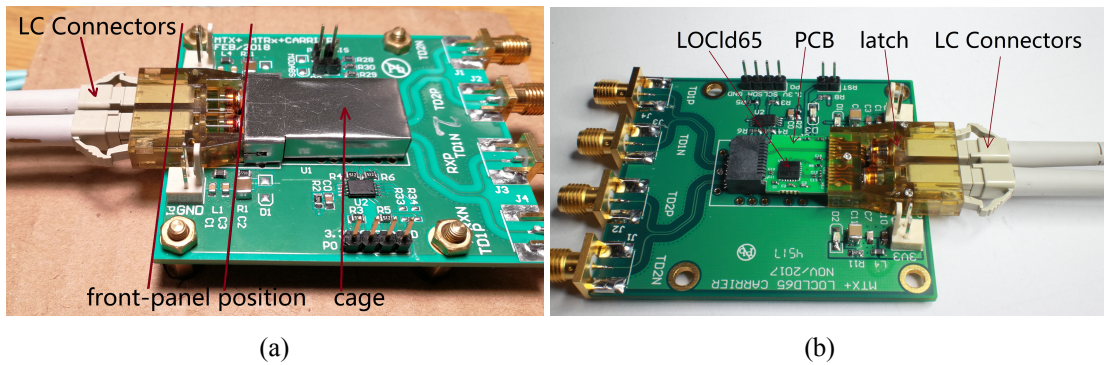

(a)                  (b)

**Figure 4.** Pictures of panel-mount (a) and board-mount (b) MTx+ plugged in the motherboard.

## 3. Test Results

A prototype MTx+ and a prototype MTRx+ were tested. The test setup is shown in Figure 5(a). A pattern generator (Model No. SDG 12070 produced by Picosecond Pulse Lab) generated a pseudorandom binary sequence ($2^7$-1) signal and fed the MTx+/MTRx+ under test. The optical output of the MTx+/MTRx+ module went to a sampling oscilloscope (Model No. TDS8000B with an optical module 80C08C produced by Tektronix) through a 2-meter optical fiber to observe eye diagrams. For MTRx+, the optical output of the transmitter was looped back to the input of the receiver and the electrical output of the receiver was sent to a real-time oscilloscope (Model DSA70804B with a differential probe P7380SMA produced by Tektronix). The pattern generator also provided a trigger clock signal for the oscilloscopes. A picture of the test setup is shown in figure 5(b).

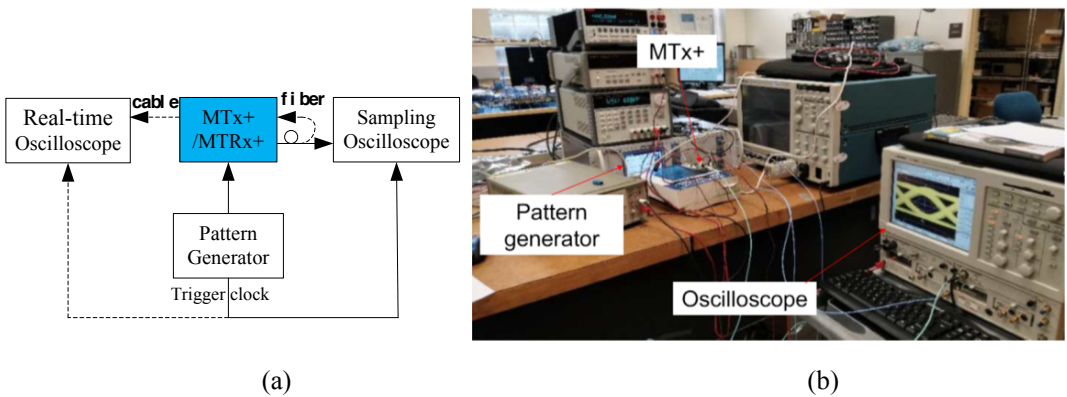

(a)                  (b)

**Figure 5.** Block diagram (a) and a picture (b) of the test setup.

An eye diagram of MTx+ operating at 14 Gbps is shown in Figure 6(a) when the VCSEL voltage is 3.3 V. The eye diagram passed the eye mask of the 10-Gbps Fiber Channel standard. The horizontal scale of the oscilloscope was adjusted to match the eye mask built in the oscilloscope. The power consumption is 68.3 mW/channel (the VCSEL included). When 3.3 V is not available, a 2.5-V voltage can be applied to MTX+ as the VCSEL voltage. With the VCSEL voltage of 2.5V, the power consumption is 62.1 mW/channel. An eye diagram of MTRx+ is shown in Figure 6(b). The receiver of MTRx+ operated at 4.8 Gbps, the same data rate as VTRx. The power consumption of the present prototype receiver was about 120 mW.

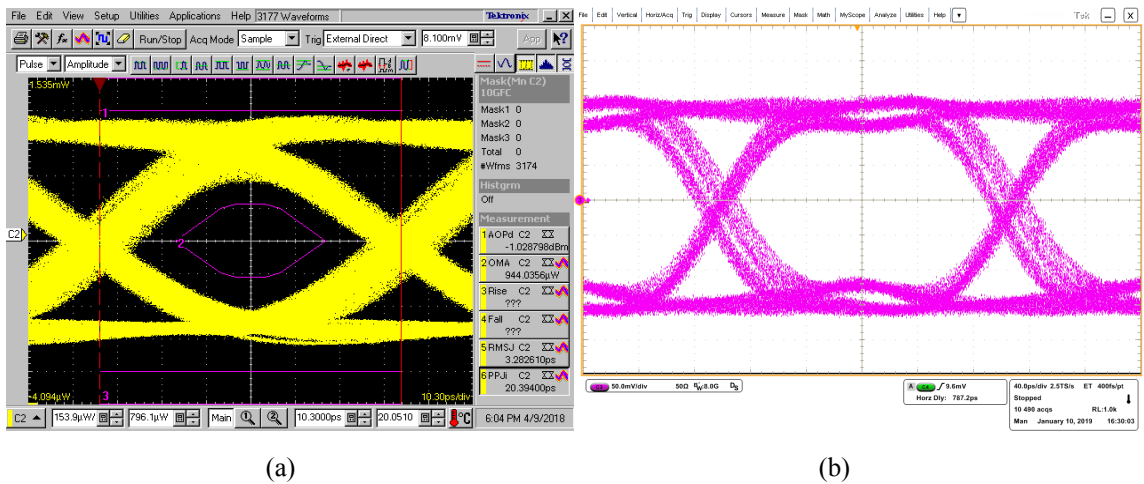

(a) (b)

**Figure 6.** Eye diagrams of MTx+ (a) and MTRx+ (b).

An MTx+ module was irradiated in x-rays with the maximum energy of 160 keV and the dose rate of 1.3 Gy (SiO$_2$)/s. The setup is the same as that shown in Figure 5(a). An x-ray irradiator (Model No. XR-160 produced by Precision X-Ray, Inc.) was used. During the test, the MTx+ module was exposed in x-rays, while all other components were either protected from the x-rays or outside of the irradiator. The data rate was 14 Gbps and the VCSEL voltage was 3.3 V. At the end of the 65-minute test, we accumulated a total ionizing dose (TID) of 4.9 kGy (SiO$_2$). The MTx+ module was functional throughout the test. The power current of 1.2V, the bias current of 3.3V, the optical modulation amplitude, and the average optical power decreased 1.0%, 2.3 %, 2.4%, and 3.6%, respectively. No significant change was observed in rise time, fall time, and jitter. An SEU test will be conducted in the future.

## 4. Conclusion

A dual-channel optical transmitter (MTx+) and an optical transceiver (MTRx+) have been developed for the front-end readout electronics of high-energy physics experiments. MTx+ utilizes two TOSAs and MTRx+ does a TOSA and a ROSA. Both MTx+ and MTRx+ receive multimode fibers with standard LC connectors as the optical interface and can be panel or board mounted to a motherboard with a standard SFP+ connector as the electrical interface. MTx+ and MTRx+ employ a dual-channel VCSEL driver ASIC LOCld65, which brings the transmitting data rate up to 14 Gbps per channel. MTx+ and MTRx+ have been tested to survive 4.9 kGy ($SiO_2$).

## Acknowledgments

The coupling of the TOSA/ROSA with fibers in MTx/MTRx was inspired by a presentation of Dr. Csaba Soos (CERN) in a Versatile Link collaboration meeting. The coupling used in MTx+/MTRx+ is a further development of MTx/MTRx. The authors would like to express their deepest appreciation for Csaba Soos and colleagues in the Versatile Link collaboration for this inspiration. This project is supported by the NSF and the DOE Office of Science, SMU's Dedman Dean's Research Council Grant, and the National Science Council in Taiwan.

## References

[1] ATLAS Collaboration, *ATLAS Phase-II Upgrade Scoping Document*, CERN-LHCC-2015-020; LHCC-G-166, September 25, 2015.

[2] L. Amaral et al., *The versatile link, a common project for super-LHC*, 2009 *JINST* 4 P12003.

[3] X. Zhao et al., *Mid-board miniature dual channel optical transmitter MTx and transceiver MTRx*, 2016 *JINST* 11 C03054.

[4] ATLAS collaboration, *ATLAS Liquid Argon Calorimeter Phase-I upgrade technical design report*, CERN-LHCC-2013-017, ATLAS-TDR-022, Dec. 2013.

[5] H. Xu, *The trigger readout electronics for the Phase-I upgrade of the ATLAS Liquid Argon calorimeters*, 2017 *JINST* 12 C03073.

[6] J. Troska et al., *The VTRx+, an Optical Link Module for Data Transmission at HL-LHC*, PoS(TWEPP-17)048.

[7] T. Zhang et al, *GBLD10+: a compact low-power 10 Gb/s VCSEL driver*, 2016 *JINST* 11 C01015.

[8] Z. Zeng et al., *LDQ10: a compact ultra low-power radiation-hard 4 x 10 Gb/s driver array*, 2017 *JINST* 12 P02020.

[9] Z. Zeng et al., *A Compact Low-Power Driver Array for VCSELs in 65-nm CMOS Technology*, *IEEE T. Nuclear Science*, vol. 64, N0. 6, Jun. 2017, p. 1599-1604.

[10] X. Li et al., Optical Data Transmission ASICs for the High-Luminosity LHC (HL-LHC) Experiments, 2014 *JINST* 9 C03007.

[11] W. Zhou et al., LOCld65, *A Dual-Channel VCSEL Driver ASIC For Detector Front-End Readout*, Presented at the Woodlands21st IEEE Real Time Conference, Colonial Williamsburg, VA, USA, Jun 9-15 2018, arXiv:1806.08851.

[12] F.X. Chang et al., Aging and Environmental Tolerance of an Optical Transmitter for the ATLAS Phase-I Upgrade at the LHC, 2015 *NIM* A831 349-354.

[13] M. Menouni et al., *The GBTIA, a 5 Gbit/s radiation-hard optical receiver for the SLHC upgrades*, presented at the Topical Workshop on Electronics for Particle Physics (TWEPP 2009), Paris, France, September 21-25, 2009.